\documentstyle[11pt]{article}

\newcommand{\eeq}{\end{equation}}

\newcommand{\vs}[1]{\rule[- #1 mm]{0mm}{#1 mm}}
\newcommand{\beq}{\begin{equation}}
\newcommand{\beql}{\begin{eqnarray}}
\newcommand{\eeql}{\end{eqnarray}}

\newcommand{\lam}{\lambda}

\newcommand{\sect}[1]{\setcounter{equation}{0}\section{#1}}

\begin{document}

\begin{titlepage}

\rightline{\Large {RRI-97-04}}

\vs{10}

\begin{center}

{\LARGE {\bf Orthogonal Polynomials and Exact\\[.5cm]
             Correlation Functions for Two Cut\\[.5cm]
             Random Matrix Models}}\\[2cm]

{\Large Nivedita Deo}\\[.5cm]
{\em Raman Research Institute\\
Bangalore 560080, India}\\
{ndeo@rri.ernet.in}\\

\end{center} 

\vs{10}

\centerline{ {\bf Abstract}}
Exact eigenvalue correlation functions are computed for
large $N$ hermitian one-matrix models with eigenvalues distributed
in two symmetric cuts. An asymptotic form for orthogonal polynomials 
for arbitrary polynomial potentials
that support a $Z_2$ symmetric distribution is obtained. This
results in an exact explicit expression for the kernel at large $N$ which
determines all eigenvalue correlators. The oscillating and smooth
parts of the two point correlator are extracted and the universality
of local fine grained and smoothed global correlators is established.
\end{titlepage}

\renewcommand{\thefootnote}{\arabic{footnote}}
\setcounter{footnote}{0}

\sect{Introduction}\label{intro}

\indent
Eigenvalue correlators in large random matrix models have
been of interest recently (see
\cite{BZ93, SLA93,NS93,B93,BZ94,P95,DJS95,D96,J96} and references
therein). The recent interest is in part inspired by the burst of 
activity in the field of quantum chaos and mesoscopic systems. 

Correlators in matrix models have been calculated in various 
limits by different methods. At short separations of the order of the
eigenvalue spacing the correlators show oscillatory behaviour.
Correlators that reproduce this behaviour are referred to as
``fine grained". If these oscillations are averaged over, one gets
the so called ``smoothed" correlators. For some applications,
one needs the correlation functions over the entire eigenvalue
distribution, near the centre as well as the edges (global 
correlators) while for others only in a small region
near the centre or the edge of the eigenvalue distribution 
(local correlators). The complete information is contained
in the global fine grained correlator, from which all others
can be obtained as special limiting cases. 

The global fine grained 
correlator has been computed for the hermitian one matrix model
\cite{BZ93}\ and the hermitian two matrix model \cite{BDZ95}\ 
using the method of orthogonal polynomials. These studies, however, 
consider only those cases where the eigenvalue distribution
lies in a single cut. The present paper is concerned with correlation 
functions for hermitian matrix models in which eigenvalues lie
in two cuts or bands.

Random matrix models with eigenvalues that have support in more
than one cut have been studied in the context
of 2-d quantum gravity and symmetry breaking solutions
(\cite{BDJT93} and references therein),
two-dimensional QCD\cite{JV96JSV96,JNZ96} and  
statistical physics \cite{CKPR95}. 
They may also be applicable in studies of certain 
condensed matter systems with eigenvalue distributions with gaps.
In the present paper the global
fine grained correlators for matrix models with eigenvalue support
in two symmetric cuts are obtained. These expressions therefore contain
detailed information about how the correlators vary over
distances of the eigenvalue spacing, from which the smoothed
correlators can be extracted by a suitable averaging procedure.
They are also valid over the entire eigenvalue distribution,
near the centre as well as the edge of the cuts.

In order to compute the global fine grained correlators,
an asymptotic form for orthogonal polynomials is derived, which is
valid for arbitrary potentials that support a $Z_2$ symmetric
eigenvalue distribution.
This form allows a determination
of an explicit expression for the eigenvalue kernel through the
Christoffel-Darboux formula
suitably modified for the two cut situation. The kernel then
determines all the global fine grained $n$-point correlation functions
of the eigenvalue density.

The two point function is explicitly
discussed in various limits. The kernel as well as the global
fine grained correlator in general depend upon the coupling
constants in the matrix model potential. However, it is
shown that in two limits the two point function is universal,
i.e., independent of the form of the potential. These two limits
are (i) fine grained local correlators away from the edge of
the cuts, where the Dyson form is reproduced, and (ii) global
smoothed correlators.

This generalizes the work of
\cite{BZ93}\ to the two cut matrix models. When the coupling
constants of the matrix model change such that the eigenvalue
support moves from two cuts to one cut, the single-cut results
are obtained. Thus the present work can be used to study this
phase transition in the coupling constant space of matrix models
at the level of fine grained correlators, generalizing earlier
studies at the level of smoothed correlators \cite{DS90,DDJT90}.

The paper is organized as follows. In the second section 
after establishing notations and conventions, 
the asymptotic form for the orthogonal polynomials for
two symmetric cuts is presented as an
ansatz. In the third and fourth sections the recurrence relation 
and the orthogonality condition are used to determine the
unknown functions in the ansatz.
The fifth section derives the Christoffel-Darboux
formula for the kernel $K(\mu,\nu)$ in a form suitable for
two cuts and an explicit expression for
kernel is presented. The sixth section discusses the two point 
correlation function in its full fine grained global form and
establishes its universality
in various limits (fine grained local, smoothed global). 
The seventh section contains concluding remarks.

\sect{Notation, Conventions and Asymptotic Ansatz} \label{notation}

\indent

The partition function is defined by
\beq
Z \ \equiv \
\int dM \ \mbox{exp}(-N \ \mbox{Tr} V(M)) \ , \label{Z}
\eeq
where $M$ is an $N\times N$ hermitian matrix and   
$V(M) \ \equiv \ \sum_{n=1}^\infty \frac{g_n}{n}M^n \ $ is a 
real polynomial in $M$ (the same notation 
as ref. \cite{BDJT93} is used).
$P_n(\lam)$ is a set of polynomials that are orthogonal
with respect to the measure defined by the potential $V$, 
$\int_{-\infty}^{\infty} d\lam P_n (\lam) P_m (\lam) 
{exp}(-N \ \mbox{Tr} V(\lam))=h_n \delta_{nm}$,
and are normalized such that $P_n(\lam)={\lam}^n+c_1 {\lam}^{n-1}+....$, 
$P_0 (\lam)=1$. These properties determine $P_n(\lam)$ uniquely for
every $V$.
The orthogonal polynomials satisfy the
recursion relation
$\lam P_n(\lam)=P_{n+1}(\lam)+S_n P_n (\lam) +R_n P_{n-1} (\lam)$
where the recursion coefficients $R_n$, $S_n$ are $V$ dependent.
$R_n = h_n/h_{n-1}$ and
for $Z_2$-symmetric potentials ($V(-\lam)=V(\lam)$) $S_n=0$. 
It is convenient to define an orthonormal set of polynomials 
$\psi_n (\lam)={P_n(\lam) \over \sqrt{h_n}}{exp}(-{N\over 2} 
\ \mbox{Tr} V(\lam))$. Then
\beq
\int d\lam \psi_n (\lam) \psi_m (\lam) 
= \delta_{nm},
\label{psipsi}
\eeq
and
\beq
\lam \psi_n(\lam)=\sqrt{R_{n+1}} \psi_{n+1}(\lam)+
\sqrt{R_n} \psi_{n-1} (\lam).
\label{xpsi}
\eeq
The kernel $K(\mu,\nu)$ is defined as
\beq
K(\mu,\nu)= {1\over N} \sum_{i=0}^{N-1} \psi_i(\mu)\psi_i(\nu).
\eeq
The Christoffel-Darboux formula expresses this in terms of just
$\psi_n$ with $n$ close to $N$:
\beq
K(\mu,\nu)={\sqrt{R_N}\over N} {[\psi_N(\mu)\psi_{N-1}(\nu)-\psi_{N-1}(\mu)
\psi_N(\nu)]\over (\mu-\nu)}.
\label{CD}
\eeq
The kernel determines all eigenvalue correlators. Defining\break
$\hat{\rho}(\mu) = {1\over N} Tr \delta (\mu-M)
= {1\over N} \sum_{i=1}^N \delta (\mu - \lam_i)$,
where $\lam_i$ are the eigenvalues of $M$,
it follows that
\beql
\rho(\mu) & \equiv & \langle \hat{\rho}(\mu) \rangle
= K(\mu, \mu)\nonumber\\
\rho_c(\mu,\nu) & \equiv &
\langle \hat{\rho}(\mu) \hat{\rho}(\nu)\rangle 
- \langle \hat{\rho}(\mu)\rangle \langle \hat{\rho}(\nu) 
\rangle
= -[K(\mu, \nu)]^2\nonumber\\
\rho_c(\lam,\mu,\nu) & \equiv &
\langle \hat{\rho}(\lam)
\hat{\rho}(\mu) \hat{\rho}(\nu)\rangle_c  
=  2 K(\lam,\mu) K(\mu,\nu) K(\nu,\lam)
\eeql
where the subscript $c$ stands for connected correlator and the expectation
value of any function $f(M)$ is given by $\langle f \rangle =
{1\over Z}\int dM e^{-NTr V(M)} f(M)$. 

When the potential is $Z_2$ symmetric and
has two wells
the eigenvalue density $\rho(\lam)$ can have support in
two cuts located symmetrically about the origin,
$\lam \in (a,b) \cup (-b,-a)$, $a \geq 0, b > a$. One is then
interested in the kernel and correlation functions where
$\mu, \nu$, etc. lie in the cuts. The Christoffel-Darboux formula
implies that we need the orthogonal polynomials for $n$ close to
$N$.

For $\lam$ lying in the two
cuts and for $N-n \sim O(1)$, $N$ large,
we make the ansatz that the orthogonal
polynomials can be approximated by 
\beq
\psi_{n} (\lam)={1\over {\sqrt{f}}}
[\cos (N\zeta-(N-n)\phi+\chi+(-1)^n \eta)+O({1\over N})],
\label{psin}
\eeq
where $f,\zeta,\phi,\chi$ and $\eta$ are functions of $\lam$,
and that $\psi_n$ is damped out outside the cuts. We then show
that 
\beq
f(\lam)={\pi\over {2 \lam}}{(b^2-a^2)\over 2}\sin 2\phi (\lam)
\label{f}
\eeq
from the orthogonality condition satisfied by the orthogonal
polynomials and is derived in section \ref{orth}.
$\zeta (\lam)$ is fixed to be 
\beq
\zeta^{\prime} (\lam)=-\pi \rho (\lam)
\label{zeta}
\eeq
from the relation
$K(\lam,\lam)=\rho(\lam)$ and is derived in section \ref{kern}.
$\phi (\lam)$ and $\eta (\lam)$ are determined from
the recurrence relation satisfied by the orthogonal polynomials 
( section \ref{rec} )
\beq
\cos 2\phi (\lam)=
{\lam^2-{(a^2+b^2)\over 2}\over {(b^2-a^2)\over 2}},
\label{phi}
\eeq
\beql
\cos 2\eta (\lam)& = &
b{\cos \phi (\lam) \over \lam},\nonumber\\
\sin 2\eta (\lam)& = &
a {\sin \phi (\lam) \over \lam}.
\label{eta}
\eeql
This leaves the function $\chi (\lam)$ undetermined and we
conjecture that it has the form ${1\over 2} \phi(\lam)
-{\pi\over 4}$ for all potentials as in the single cut
case ref. (\cite{BZ93,E93}).

Example for $g_2 < 0, g_4 > 0$, all other $g_i = 0$,
and in the two cut phase $g_2^2>4 g_4$, 
the end points are given by
$a^2=-2\sqrt{1\over g_4}-{g_2\over g_4}$
and $b^2= 2\sqrt{1\over g_4}-{g_2\over g_4}$. The density of eigenvalues is
$\rho(\lam)={g_4\over 2\pi} \lam \sqrt{(\lam^2-a^2)(b^2-\lam^2)}$
for $\lam \in (-b,-a) \cup (a,b)$ and zero elsewhere.
Hence $\zeta(\lam)=\phi(\lam)-{\sin 4\phi(\lam)\over 4}$.
This determines the asymptotic form of $\psi_n(\lam)$ completely.

Note that $f, \phi, \eta$ and $\chi$ are universal functions, i.e,
their functional form is independent of the potential $V$, the only
dependence on $V$ enters through the endpoints of the cuts $a$ and $b$.
$\zeta$ is non-universal since the eigenvalue density depends in general 
on the detailed form of $V$. 

\sect{Recurrence Relation} \label{rec}

We will now use Eq. (\ref{xpsi}) to determine $\phi$ and $\eta$.
Multiplying Eq. (\ref{xpsi}) by $\lam$ and using Eq. (\ref{xpsi})
once again we get
\beql
\lam^2 \psi_n (\lam) &=& \sqrt{R_{n+1}}\sqrt{R_{n+2}}\psi_{n+2}(\lam)
+R_{n+1}\psi_n (\lam) +R_n \psi_n (\lam) \nonumber \\
& + & \sqrt{R_n}\sqrt{R_{n-1}} \psi_{n-2} (\lam).
\label{x2psi}
\eeql
When $V$ has two symmetric wells, it is known that 
$R_n=A_n$ for $n$ even and $R_n=B_n$ for $n$ odd, where in the
large N limit $A_n$ and $B_n$ are approximated by two
continuous functions $A(x)$, $B(x)$ with $x={n\over N}$
\cite{CMM86,DDJT90}. Thus, for $n$ even
\beql
\lam^2 \psi_{n} (\lam) & = & 
\sqrt{B_{n+1}}\sqrt{A_{n+2}}\psi_{n+2}(\lam)
+B_{n+1}\psi_{n}(\lam)\nonumber\\
& + & A_{n}\psi_{n}(\lam)+\sqrt{A_{n}}\sqrt{B_{n-1}}
\psi_{n-2}(\lam).
\label{x2abpsi}
\eeql
On using the asymptotic ansatz for even $n$ 
i.e., substituting Eq. (\ref{psin})
in Eq. (\ref{x2abpsi}), and replacing $A_n=A(x={n\over N})=
A(x=1-{N-n\over N})=A(1)+O({1\over N})\approx A(1)\equiv A$
(and similarly $B_n\rightarrow B(1)\equiv B)$ we get 
\beq
{\lambda^2-(A+B)\over 2\sqrt{AB}} = \cos 2\phi (\lam).
\label{cos2phi}
\eeq
It is known that for $Z_2$ symmetric two cut models the
end points $a$ and $b$ are related to $A$ and $B$ by \cite{DDJT90}
\beql
A+B &=& {a^2+b^2 \over 2},\nonumber\\
2\sqrt{AB} &=& {b^2-a^2 \over 2}.
\eeql
This yields Eq. (\ref{phi}).

Next consider the recurrence relations ($n$ even)
\beql
\lam \psi_{n+1} (\lam) &=& \sqrt{R_{n+2}}\psi_{n+2} (\lam)
+ \sqrt{R_{n+1}} \psi_n (\lam)\nonumber\\
\lam \psi_{n-1} (\lam) &=& \sqrt{R_n} \psi_n (\lam)
+ \sqrt{R_{n-1}} \psi_{n-2} (\lam).
\label{psin1}
\eeql
On substituting Eq. (\ref{psin}) for $\psi_n$,
$\psi_{n\pm 1}$, $\psi_{n\pm 2}$ in
Eq. (\ref{psin1}) it is easy to see that
\beql
\sin 2\eta (\lam) &=& (-\sqrt{A}+\sqrt{B})
{\sin \phi (\lam) \over \lam}\nonumber\\
&=& a {\sin \phi (\lam) \over \lam}
\eeql
and
\beql
\cos 2\eta (\lam) &=& (\sqrt{A}+\sqrt{B})
{\cos \phi (\lam) \over \lam}\nonumber\\
&=& b {\cos \phi (\lam) \over \lam}.
\eeql
This determines Eqs. (\ref{phi}) and (\ref{eta}). 

\sect{Orthogonality} \label{orth}

Let us check for orthonormality, Eq. (\ref{psipsi}),
for all $n, m$. The ansatz assumes that for large $N$
$\psi_n (\lam)$ is damped out
sufficiently fast beyond the end of cuts so that integration
can be restricted to the cuts
\beq
(\int_{-b}^{-a}+\int_{a}^{b}) d\lam \psi_n (\lam)
\psi_m (\lam)=\delta_{nm}.
\eeq
The integrand is
\beql
\psi_n \psi_m &=& {1\over f} \cos \alpha \cos \beta\nonumber\\
&=&{1\over 2f} [\cos(\alpha + \beta) + \cos (\alpha - \beta)]
\eeql
where $(\alpha+\beta)=(2N\zeta-(2N-n-m)\phi+2\chi+(-1)^n\eta)$
and $(\alpha-\beta)=((n-m)\phi+((-1)^n-(-1)^m)\eta)$. The
$\cos(\alpha+\beta)$ term is zero upon integration because 
$\cos(2N\zeta-(2N-n-m)\phi+2\chi+(-1)^n\eta)$ oscillates rapidly
and averages to zero. Thus
\beq
(\psi_n,\psi_m)= (\int_{-b}^{-a}+\int_{a}^{b}) {d \lam\over
2 f} \cos (\alpha-\beta)
\eeq
Note that $(-1)^n-(-1)^m=0$ if $n-m$ is even and $(-1)^n-(-1)^m=2(-1)^n$
if $n-m$ is odd. For $n-m$ even
\beq
(\psi_n,\psi_m)=(\int_{-b}^{-a}+\int_{a}^{b}) {d\lam\over 2f}
\cos (n-m)\phi.
\label{psieven}
\eeq
In particular for $n-m=0$
\beq
(\psi_n,\psi_n)=(\int_{-b}^{-a}+\int_{a}^{b}) {d\lam\over 2f}.
\label{psinpsin}
\eeq
We will show that if
\beq
{{d \lam} \over {d \phi}} = c f(\lam),
\label{dldpfc}
\eeq
then orthogonality follows.
On taking the derivative of Eq. (\ref{phi}) one gets
\beq
{{d\lambda} \over {d\phi}} 
={-1\over \lam}{(b^2-a^2)\over 2} \sin 2\phi 
\label{dldp}
\eeq
Thus
\beql
f(\lam)
= {-1\over c}{(b^2-a^2)\over 2}{\sin 2\phi(\lam)\over \lam}.
\label{fc}
\eeql
Note from Eq. (\ref{psinpsin}) that for $(\psi_n,\psi_n)=1$, 
$f(\lam)$ must be an even
function of $\lam$. This implies that $\sin 2\phi$ must be
an odd function of $\lam$. Since $\cos 2\phi=-1$ at $\lam=\pm a$,
and 1 at $\lam=\pm b$, this is achieved by taking $2\phi=2p\pi,
(2p-1)\pi, (2p-3)\pi, (2p-4)\pi$ at $\lam=-b,-a,a,b$ respectively;
with $p$ any integer.
With this choice $\sin 2\phi$ is odd because
$2\phi$ is in the first two quadrants for $\lam$ in $(a,b)$
and the third and fourth quadrants for $\lam$ in $(-b,-a)$.
In particular this choice also implies that
$\cos \phi$ is an even function of $\lam$ and $\sin \phi$ is an
odd function. Substituting Eq. (\ref{fc}) into
Eq. (\ref{psinpsin}) implies that $c={-2\over \pi}$.
Using $c={-2\over \pi}$ in Eq. (\ref{fc}) gives Eq. (\ref{f}).

Now consider $n-m \neq 0$ even. Using
Eq. (\ref{dldpfc}) in Eq. (\ref{psieven}), we find
\beql
(\psi_n,\psi_m) &=&
c(\int_{\phi(-b)}^{\phi(-a)}+\int_{\phi(a)}^{\phi(b)})
d\phi \cos (n-m)\phi\nonumber\\
&=&0.
\eeql
Though the integrand is an even function of $\lam$ the
full integral vanishes as each integral separately vanishes.

Now consider $n-m$ odd
\beq
(\psi_n,\psi_m)=(\int_{-b}^{-a}+\int_a^b){d\lam\over 2f}
\cos (\alpha-\beta)
\eeq
where $\alpha-\beta=(n-m)\phi+2(-1)^n\eta$ for $n-m$ odd.
Expanding $\cos (\alpha-\beta)$ we get
\beq
(\psi_n,\psi_m) = (\int_{-b}^{-a}+\int_a^b) {d\lam\over 2f}
[\cos (n-m)\phi \cos 2\eta-(-1)^n \sin (n-m)\phi \sin 2\eta]
\label{odd}
\eeq
Consider the first term in the bracket $[\,\,]$. Since $\cos\phi$ 
is an even function of $\lam$, $\cos (n-m)\phi$ is even and 
$\cos 2\eta$ is odd (see Eq. (\ref{eta})). Thus the integrand is
odd and gives zero upon integrating over $(-b,-a) \cup (a,b)$. Similarly
the second term in the $[ \,\,]$ is also odd, because $\sin \phi$ is odd 
which implies $\sin (n-m)\phi$ is odd and $\sin 2\eta$ is even from
Eq. (\ref{eta}). Thus $(\psi_n,\psi_m)=0$ for $n-m$ odd.

This completes the proof of the orthonormality of $\psi_n (\lam)$.  

\sect{Kernel}\label{kern}

We will obtain an expression for $K(\mu,\nu)$ analogous to
the Christoffel-Darboux formula from Eq. (\ref{CD}), which is
more convenient for the present two cut structure.
Multiplying and dividing Eq. (\ref{CD}) by $\mu^2-\nu^2$, the kernel
for $N=2P$ even is
\beql
K (\mu,\nu)&=&{1\over 2P} [\sum_{n=0}^{(P-1)} 
{{\mu^2\psi_{2n}(\mu)\psi_{2n}(\nu)
-\psi_{2n}(\mu)\nu^2\psi_{2n}(\nu)}\over {(\mu^2-\nu^2)}}\nonumber\\
&+&\sum_{n=0}^{(P-1)} {{\mu^2 \psi_{2n+1}(\mu) \psi_{2n+1}(\nu)
-\psi_{2n+1} (\mu) \nu^2 \psi_{2n+1} (\nu)}\over {(\mu^2-\nu^2)}}]
\eeql
Using the recurrence relationship for the even and odd
orthogonal polynomials and with some algebra we get 
\beql
&&K (\mu,\nu)={1\over 2P} {1\over (\mu^2-\nu^2)}[\sqrt{A_{2P}B_{2P+1}}
(\psi_{2P+1}(\mu)\psi_{2P-1}(\nu)\nonumber\\
&-&\psi_{2P-1}(\mu)\psi_{2P+1}(\nu))
+\sqrt{B_{2P-1}A_{2P}}(\psi_{2P}(\mu)\psi_{2P-2}(\nu)\nonumber\\
&-&\psi_{2P-2}(\mu)\psi_{2P}(\nu))]
\label{keven}
\eeql

Following a similar procedure for $N=2P+1$ odd the kernel $K (\mu,\nu)$ is 
\beql
K (\mu,\nu)&=&{1\over {(2P+1)}} {1\over (\mu^2-\nu^2)}[\sqrt{A_{2P}B_{2P+1}}
(\psi_{2P+1}(\mu)\psi_{2P-1}(\nu)\nonumber\\
&-&\psi_{2P-1}(\mu)\psi_{2P+1}(\nu))
+\sqrt{B_{2P+1}A_{2P}}(\psi_{2P+2}(\mu)\psi_{2P}(\nu)\nonumber\\
&-&\psi_{2P}(\mu)\psi_{2P+2}(\nu))]
\label{kodd}
\eeql

The kernel in this form is more useful than the standard form for
the symmetric two cut hermitian matrix model due to the explicit
form of $f$ and $\phi$ (Eqs. (\ref{f}) and  (\ref{phi})) in which 
$2\phi$ appears rather than $\phi$.

A convenient form for the kernel with two cuts as derived above is
(for $N$ even)
\beql
& & K (\mu,\nu) = K^{(o)} (\mu,\nu) + K^{(e)} (\mu,\nu)\nonumber\\
&=& {1\over N} {1\over (\mu^2-\nu^2)}
[\sqrt{A_{N} B_{N+1}} (\psi_{N+1}(\mu) \psi_{N-1}(\nu)
-\psi_{N-1}(\mu)\psi_{N+1}(\nu))]\nonumber\\
&+&{1\over N} {1\over (\mu^2-\nu^2)}
[\sqrt{A_{N} B_{N-1}} (\psi_{N}(\mu) \psi_{N-2}(\nu)
-\psi_{N-2}(\mu)\psi_{N}(\nu))].
\eeql
Using the asymptotic ansatz and doing some simple trigonometry with
$\psi_{N+1} (\mu) = {1\over \sqrt{f(\mu)}} \cos (N\zeta +\phi+\chi
-\eta)(\mu)$,
$\psi_{N-1} (\nu) = {1\over \sqrt{f(\nu)}} \cos (N\zeta -\phi+\chi
-\eta)(\nu)$,
$\psi_{N} (\mu) = {1\over \sqrt{f(\mu)}} \cos (N\zeta + \chi
+\eta)(\mu)$ and
$\psi_{N-2} (\nu) = {1\over \sqrt{f(\nu)}} \cos (N\zeta -2\phi+\chi
+\eta)(\nu)$
we get the following
\beql
K^{(o)} (\mu,\nu) &=& {-1\over 2N} {\sqrt{A_{N} B_{N+1}}\over {(\mu^2-\nu^2)
\sqrt{f(\mu)f(\nu)}}}\nonumber \\
& [ & (\cos 2\phi (\mu) - \cos 2\phi (\nu))
[\cos N (h^{(o)}(\mu)+h^{(o)}(\nu))\nonumber\\
&+&\cos N (h^{(o)}(\mu)-h^{(o)}(\nu))]
\nonumber\\
&+& (\sin 2\phi(\mu)-\sin 2\phi(\nu))\sin N(h^{(o)}(\mu)+h^{(o)}(\nu))
\nonumber\\
&+&(\sin 2\phi(\mu)+\sin 2\phi(\nu))\sin N(h^{(o)}(\mu)-h^{(o)}(\nu))]
\label{ko}
\eeql
where $Nh^{(o)}(\mu)=(N\zeta+\chi+\phi-\eta)(\mu)$ and
\beql
K^{(e)} (\mu,\nu) &=& {-1\over 2N} {\sqrt{A_{N} B_{N-1}}\over {(\mu^2-\nu^2)
\sqrt{f(\mu)f(\nu)}}} \nonumber \\
& [ & (\cos 2\phi (\mu) - \cos 2\phi (\nu))
[\cos N (h^{(e)}(\mu)+h^{(e)}(\nu))\nonumber\\
&+&\cos N (h^{(e)}(\mu)-h^{(e)}(\nu))]
\nonumber\\
&+& (\sin 2\phi(\mu)-\sin 2\phi(\nu))\sin N(h^{(e)}(\mu)+h^{(e)}(\nu))
\nonumber\\
&+&(\sin 2\phi(\mu)+\sin 2\phi(\nu))\sin N(h^{(e)}(\mu)-h^{(e)}(\nu))]
\label{ke}
\eeql
where $Nh^{(e)}(\mu)=(N\zeta+\chi+\eta)(\mu)$.
Using the fact that $\lim_{\nu \rightarrow \mu} K(\mu,\nu)=\rho(\mu)$ one finds that
$\zeta^{\prime} (\mu) = -\pi \rho (\mu)$. In more detail, the first term in 
$K^{(o)} (\mu,\nu)$ and $K^{(e)} (\mu,\nu)$ is of order $O({1\over N})$
while the second and third terms give a term of order $O(1)$ which 
arises due to a Taylor expansion about $\nu=\mu+\delta$. The $O(1)$ terms
each give a factor ${-1\over 2\pi} h^{\prime (e)} (\mu)$ and
${-1\over 2\pi} h^{\prime (o)} (\mu)$ on taking the 
$\nu \rightarrow \mu$ limit. These terms then combine to give
${-1\over \pi}\zeta^{\prime} (\mu)$. 
This determines the 
unknown function $\zeta (\mu)$ in the asymptotic 
ansatz $\psi_n (\mu)$ in terms of the density of eigenvalues,
and yields Eq. (\ref{zeta}). 

Eqs. (\ref{ko}) and (\ref{ke}) together with $K=K^{(o)}+K^{(e)}$,
and Eqs. (\ref{f}-\ref{eta}) describe the full kernel
function for the hermitian one matrix model with
eigenvalues distributed in two
disjoint cuts. The result is valid for any polynomial potential
which can support two symmetric cuts. When $a=0$ the two cuts
in the density of eigenvalues merge into one, there is a phase
transition from a
double band to a single band. The two-cut ansatz, Eq. (\ref{psin}) for 
$\psi_n(\lam)$, changes to the one-cut ansatz of ref. \cite{BZ93} Eq. (2.6)
at $a=0$. Thus the exact kernel derived above
contains all the information of phase transitions and universality for both 
the fine and coarse grained correlators. 

\sect{Correlation Functions}\label{corr}

The full two-point connected correlation function for the
two-cut hermitian matrix model is
\beql
\rho_c (\mu,\nu)&=&- [K(\mu,\nu)]^2= -( [K^{(o)}(\mu,\nu)]^2
+ [K^{(e)} (\mu,\nu)]^2\nonumber\\
&+& 2 K^{(o)} (\mu,\nu) K^{(e)} (\mu,\nu) )
\label{full}
\eeql
with $K^{(o)}(\mu,\nu)$ and $K^{(e)}(\mu,\nu)$ given by 
Eq. (\ref{ko}) and Eq. (\ref{ke}) respectively.

We now discuss the correlator in various limits.

\noindent
(a). First consider the fine grained correlator i.e. where the 
separation $\mu-\nu$ of the order of the eigenvalue spacing $\approx 
O({1\over N})$ ;
\beq
(\mu-\nu) \approx O({1\over N})\equiv 2\delta
\eeq
Then only the third terms in $K^{(o)}(\mu,\nu)$
and $K^{(e)}(\mu,\nu)$ contribute to $K (\mu,\nu)$. Thus (for $N$ even)
\beq
K (\mu,\nu)\rightarrow {{\sin[2N\pi\delta\rho(\bar{\mu})]}
\over {2N \pi \delta}}
\eeq
where $\bar{\mu}={{\mu+\nu}\over 2}$. Over the range $\delta$, 
$\rho(\bar{\mu})$ is just constant which can be scaled away.
This extends the validity of Dyson's short distance universality to eigenvalues
distributed in two disconnected cuts. The two point correlation function is 
\beq
\rho(\mu,\nu)=\rho(\mu)\rho(\nu)[1-({{\sin x} \over x})^2]
\eeq
with $x=2N\pi\delta\rho(\bar{\mu})$. 

\noindent
(b). For $\delta<<O({1\over N})$ and $x<<1$ one gets
\beq
\rho(\mu,\nu) \approx {1\over 3} \pi^2 \rho^4 (\bar{\mu})[N(\mu-\nu)]^2.
\eeq
It is evident that the exact short distance correlator has no singularity
as $\nu\rightarrow \mu$.

\noindent
(c). In the smoothing regime, i.e., for $\mu-\nu>>O({1\over N})$ we average over
$\mu,\nu$ in an interval $\Delta$ such that ${1\over N}<<\Delta<<1$.
Replacing $<\cos^2 Nh>$ and $<\sin^2 Nh>$ by a half and 
$<\cos Nh>=<\sin Nh>=0$ 
\beql
\rho^{smooth}_c (\mu,\nu)&=&-<[K (\mu,\nu)]^2>\nonumber\\
&=& -{1\over {2 N^2\pi^2}} {\mu\nu\over (\mu^2-\nu^2)^2}
{1\over \sin 2\phi(\mu) \sin 2\phi(\nu) }\nonumber\\
& [ & (1-\cos 2\phi(\mu) \cos 2\phi(\nu))(1+\cos \bar{\phi}(\mu)
\cos \bar{\phi}(\nu))\nonumber\\
&+& \sin 2\phi(\mu) \sin 2\phi(\nu)
\sin \bar{\phi}(\mu) \sin \bar{\phi}(\nu) ],
\label{rhosm}
\eeql
where $\bar{\phi}(\mu)=\phi(\mu)-2\eta(\mu)$,
is the global smoothed correlator valid all over the cuts.
It is interesting to note that even when $\mu$ and $\nu$ belong 
to the same segment the smoothed correlator has changed from
the single band universal result (ref. \cite{BZ93} Eq. (2.22)). 

This proves the universality of smoothed global two point correlators
(as the correlator depends only on the end points of the cuts
$(-b,-a)$ and $(a,b)$) for  
symmetric two-cut matrix models, generalizing the 
universality for the one cut matrix models proven
in \cite{AJM90,BZ93}. Note that the function $\chi$ 
undetermined in our approach
is absent in the universal limits $ (a), (b)$ and $(c)$.

\sect{Conclusion}\label{conc}
\indent

In this work, the orthogonal polynomial method for large $N$
hermitian matrix models
with arbitrary potentials that support eigenvalues distributed
in two symmetric cuts has been developed. An asymptotic
form for the orthogonal polynomials $\psi_n(\lam)$ valid for $n$ close to
$N$ and $\lam$ inside the cuts has been obtained. This is done by
making an ansatz in terms of unknown
functions $f, \phi, \eta, \chi$ and $\zeta$.
The first three are then constrained using
the recursion relations and orthonormality, which determine them
to be universal functions of $\lam$ independent of the form of
the potential except for the endpoints of the cuts. $\chi$, though
not determined in this approach is also conjectured to be universal.
The crucial difference between the
one-cut case and the two-cut case appears 
in the behaviour of the recursion coefficients, which
have a ``period two" character, i.e., are described by two
continuous functions in the large $N$ limit instead of one.
This modifies the functional form of the functions $f$ and $\phi$
and introduces the new nontrivial function $\eta$ in the asymptotic
ansatz that vanishes in the single cut case.

The orthogonal polynomials are then used in a Christoffel-Darboux
formula (a version of this formula especially suited for two
symmetric cuts is used) to determine the eigenvalue kernel
$K(\mu,\nu)$. The relation between this kernel and the
eigenvalue density fixes the last unknown function $\zeta$ in the ansatz.
This is the only non universal function in the asymptotic formula
for the orthogonal polynomials and hence also in the kernel and the
correlation functions.

The kernel determines global fine grained eigenvalue
$n$-point correlation functions through well known formulas.
The two point function is discussed in detail. It is shown to
be universal in two limits, the local fine grained limit and the global
smoothed limit (in the latter it depends on the potential only
through the endpoints of the cuts). It is hoped that the
above can be generalized to multi-cut as well as
complex matrix models.

\noindent
\underline{Note added}:

After this work was completed I became aware of \cite{BI97}, where the
orthogonal polynomials are obtained for the case of quartic potential
$V(M)={g_2\over 2} M^2 +{g_4\over 4} M^4$ with two cuts
by different methods. I thank Prof. E. Brezin for information about
this work. The method and results of the present paper are valid
more generally, i.e., for arbitrary $Z_2$ symmetric polynomial potentials.

I would like to thank the referee for pointing out the
following references: i) \cite{HINS97} for a renormalization group approach
to universality of smoothed correlators in multi-cut models, 
ii) \cite{JV96JSV96} for applications of two-cut models to QCD, and
iii) \cite{AA96} for computation of smoothed Green function 
for multi-cut models. 
It is not clear that the connected density-density correlator derived from 
\cite{AA96} is the same as the one presented here. This may be due to the 
different limiting procedures used (e.g. the $N\rightarrow \infty$ and
$Asym\rightarrow 0$ limits do not commute \cite{BDJT93}). I would
also like to thank G. Akemann for an email discussion on this point.

\begin{flushleft}

\underline{Acknowledgements}: 

I would like to thank E. Brezin and S. Jain for discussions. 

\end{flushleft}

\newpage


\newcommand{\NP}[3]{{\it Nucl. Phys. }{\bf B#1} (#2) #3}
\newcommand{\PL}[3]{{\it Phys. Lett. }{\bf B#1} (#2) #3}
\newcommand{\PR}[3]{{\it Phys. Rev. }{\bf #1} (#2) #3}
\newcommand{\PRL}[3]{{\it Phys. Rev. Lett. }{\bf #1} (#2) #3}
\newcommand{\IMP}[3]{{\it Int. J. Mod. Phys }{\bf #1} (#2) #3}
\newcommand{\MPL}[3]{{\it Mod. Phys. Lett. }{\bf #1} (#2) #3}
\newcommand{\JP}[3]{{\it J. Phys. }{\bf A#1} (#2) #3}

\end{document}